\documentclass[draft,envcountsame]{llncs} 
\usepackage{amssymb}

\title{%
Reduction of Algebraic Parametric Systems\\%
by Rectification of their\\%
Affine Expanded Lie Symmetries%
}
\titlerunning{Reduction of Parametric Systems by Rectification of
  Expanded Lie Symmetries}
\author{%
  Alexandre Sedoglavic}
\institute{%
  ALIEN Project, INRIA Futurs \& LIFL (CNRS, UMR 8022), \\%
  Universit\'e des Sciences et Technologie de Lille, 59655 Villeneuve
  d'Ascq, France.
}
\def\lfrac#1#2{{ #1 }/ { #2 }}
\usepackage{bbm} 
\def\caracteristic#1{\bbbone_{#1}} 
\def\invariant#1{\mathrm{#1}}
\def\invariantset{\zeta}

\def\integer{\mathbb{N}}

\DeclareMathAlphabet{\mathpzc}{OT1}{pzc}{m}{it}
\def\freeparam#1{\mathpzc{#1}}
\def\function{\mathrm{f}}

\def\polynomialfunction{\mathrm{p}}
\def\algebra{\mathrm{A}}
\def\field#1{\mathbb{#1}}
\def\basefield{\field{K}}

\def\fractionfieldelement{h}

\def\localization#1#2{{#1}_{(#2)}}
\def\multiplicativeset{\wp}
\def\flowtime{\tau}
\def\eigenvalue{\lambda}
\def\eigenmultiplicity{\alpha}
\def\principalelement{\rho}
\def\preprincipalelement{\varrho}
\def\slice{q}
\def\originalsolution{\mathcal{V}}

\def\inducedsolution{\mathcal{W}}
\def\inducedvariety{W}

\def\hyperplan{\mathcal{H}_{\InfGen{\delta}}}

\def\Liealgebra#1{\textup{LieSym}({#1})}

\def\DerModule#1#2{\textup{Der}_{#1}#2}

\def\AffDer#1#2{\textup{AffDer}_{#1}#2}
\def\InfGen#1{#1}

\def\Matrix#1{\mathcal{#1}}
\def\LinearPart#1{\Matrix{A}_{#1}}
\def\TransPart#1{\Matrix{B}_{#1}}
\def\VariableChangeMatrix{\Matrix{P}}
\def\ExponentialMap#1{e^{#1}}

\def\automorphism{\sigma}

\def\projection#1#2{\pi_{#1,#2}}

\def\constant{c}
\def\constantvector{C}
\def\variable{z}
\def\variableset{Z}
\def\variablenb{m}

\def\variablenb{m}
\def\OrdinaryDifferentialSystem{\Sigma}
\def\time{t}
\def\statevariable{x}
\def\statevariableset{X}
\def\statevariablenb{n}
\def\parameter{\theta}
\def\parameterset{\Theta}
\def\parameternb{\ell}
\def\VectorFieldSecondMember{f}
\def\VectorFieldSecondMemberset{F}
\def\statevariableindiceset{N}
\def\indicesubset{J}
\def\indicep{\imath}
\def\indices{\jmath}

\spnewtheorem*{examplecontinued}{Example~1 (continued)}{\itshape}{\rmfamily}
\spnewtheorem*{sketchofproof}{Sketch of proof}{\itshape}{\rmfamily}
\spnewtheorem{hypotheses}{Hypotheses}{\itshape}{\rmfamily}
\begin{document}
\maketitle
\begin{abstract}
  Lie group theory states that knowledge of a~$m$-parameters solvable
  group of symmetries of a system of ordinary differential equations
  allows to reduce by~$m$ the number of equations.  We apply this
  principle by finding some \emph{affine derivations} that induces
  \emph{expanded} Lie point symmetries of considered system.  By
  rewriting original problem in an invariant coordinates set for these
  symmetries, we \emph{reduce} the number of involved parameters.  We
  present an algorithm based on this standpoint whose arithmetic
  complexity is \emph{quasi-polynomial} in input's size.
\end{abstract}
\section{Introduction}
\label{sec:Introduction}
Before analysing a biological model described by an algebraic system,
it is useful to reduce the number of relevant parameters that
determine the dynamics.
\begin{example}
  \label{ex:VerhulstLinearPredation}%
  In order to give an example of such a reduction, let us consider the
  following Verhulst's logistic growth model with linear predation
  (see~\S~1.1 in~\cite{Murray2002}):
  \begin{equation}
    \label{eq:VerhulstLinearPredation}
    \dot{x} = (a-bx)x - c x, 
    \quad 
    \dot{a}=\dot{b}=\dot{c}=0,\quad \dot{t}=1,
  \end{equation}
  for which all forthcoming computations could be easily performed by
  hand. Assuming that~${a\not =c}$, one can represent the flow~$(t,x)$
  of~(\ref{eq:VerhulstLinearPredation}) using parameterization:
  \begin{equation}
    \label{eq:InvariantsVerhulstLinearPredation}
    t =  \invariant{t}/(a-c),\qquad
    x =  (a-c)\invariant{x}/b,
  \end{equation}
  where~$(\invariant{t},\invariant{x})$ is the flow of the following
  differential equation:
  \begin{equation}
    \label{eq:ReducedVerhulstLinearPredation}
    \dot{\invariant{x}} = (1-\invariant{x})\invariant{x}. 
  \end{equation}
  In this formulation of~(\ref{eq:VerhulstLinearPredation}),
  parameters~$a$ and~$c$ were lumped together into~${a-c}$ and its
  state variables~$x$ and~$t$ were \emph{nondimensionalise}.
\end{example}
Usually, presentation of this kind of simplification relies on rules
of thumbs (for example, the knowledge of units in which is expressed
the problem when dimensional analysis is used) and thus, there is---up
to our knowledge---no complexity results on these kind of reduction
methods (see~\cite{Khanin2001} and references therein).
\par
However, these reductions are generally based on the existence of Lie
point symmetries of the considered problem (for reduction based on
dimensional analysis, see the Theorem~3.22 in~\cite{Olver1993}).
\begin{examplecontinued}
  The following continuous groups of transformations:
  \begin{equation}
    \label{eq:VerhulstLinearPredationSymmetries}
    \mathcal{T}_{\lambda}:    
    \begin{array}{ccr} 
      \freeparam{t} & \rightarrow & \freeparam{t}, \\
      \freeparam{x} & \rightarrow & \freeparam{x},
    \end{array} 
    \quad
    \begin{array}{ccc}
      \freeparam{a} &\rightarrow& \freeparam{a} - \lambda, \\
      \freeparam{b} &\rightarrow& \freeparam{b}, \\
      \freeparam{c} &\rightarrow& \freeparam{c} - \lambda, 
    \end{array}
    \qquad
    \mathcal{S}_{(\mu,\nu)}:
    \begin{array}{ccr} 
      \freeparam{t} & \rightarrow & \freeparam{t}/\nu, \\
      \freeparam{x} & \rightarrow & \mu \freeparam{x},
    \end{array} 
    \quad    \begin{array}{ccc} 
      \freeparam{a} & \rightarrow & \nu \freeparam{a}, \\
      \freeparam{b} & \rightarrow & \nu \freeparam{b}/\mu, \\
      \freeparam{c} & \rightarrow & \nu \freeparam{c},
    \end{array} 
  \end{equation}
  leave invariant system~(\ref{eq:VerhulstLinearPredation}) and its
  solutions. These symmetries are called \emph{expanded} because they
  act on the expanded space of variables that includes the system
  parameters in addition to independent and dependent variables.
\end{examplecontinued}
The system~(\ref{eq:ReducedVerhulstLinearPredation}) is obtained by
`factoring out' the actions of the
symmetries~(\ref{eq:VerhulstLinearPredationSymmetries}) of the
original system~(\ref{eq:VerhulstLinearPredation}) and thus, it is
invariant under these actions.  The
relations~(\ref{eq:InvariantsVerhulstLinearPredation}) parameterize
the solutions of the original
system~(\ref{eq:VerhulstLinearPredation}) in function of the
solution~$(\invariant{t},\invariant{x})$ of invariant
system~(\ref{eq:ReducedVerhulstLinearPredation}) and of the free
parameters~$a,b,c$; they are defined by the
composition~${(\mathcal{S}_{(\freeparam{a}/\freeparam{b},\freeparam{a})}
\circ
\mathcal{T}_{\freeparam{c}})(\invariant{t},\invariant{x},a,b,c)}$.
\par\medskip\noindent The aim of this note is to show how Lie theory
unifies and extends the classical methods (exact lumping, dimensional
analysis, etc.) used to simplify algebraic (differential) systems. We
adopt a presentation based on algebraic tools closer to the actual
computations (mainly Jordan normal form and linear algebra) on which
are based our reduction process.
\subsection{Related works}
\label{sec:RelatedWorks}
The literature on
investigation of the invariants of Lie group action and their applications
to algebraic systems is far too vast to be reviewed properly here.  
The book~\cite{Gatermann2000} shows various applications of invariants
in the study of dynamical systems under a computer algebra viewpoint.
Section~4.1 for example, shows how the knowledge of invariants of a
given dynamical system could simplify further computation on it by
reducing the degree of involved polynomial expressions.
In~\cite{HubertKogan2006} authors show, given a rational group action,
how to compute a complete set of its invariants.
\par
We adopt the same general philosophy---determine some system's
symmetries and use their invariants---but our purposes is more to
reduce the number of variables involved in these expressions than
their degrees (that is for us only a byproduct).  Furthermore, while
the symmetries considered in~\cite{Gatermann2000,HubertKogan2006} are
quite general---and thus, their computation of invariants are
exclusively done using Gr\"obner bases computations---we restrict
ourself to the use of \emph{affine} Lie symmetries and thus, the
required operations are restricted to linear algebra over a number
field and univariate polynomial factorization.
\subsection{Main Steps and Tools of the Reduction Process}
The reduction process introduced in this note is based on classical
Lie theory but, to the best of our knowledge, it is not described
elsewhere in the literature.  Let us present now the tools used in
this process through our introducing example:
\begin{examplecontinued}
  Reduction of Example~\ref{ex:VerhulstLinearPredation} is done as follow
  \begin{itemize}
  \item[Step~1.] Determine affine infinitesimal generators that induce
    expanded Lie symmetries of the considered system.  If there is no
    such derivations, our reduction process stops.  In our example,
    they are:
    \begin{equation}
      \label{eq:VerhulstInfGen}
      \textstyle
      \InfGen{\delta}_{1} := \frac{\partial\hfill}{\partial a} 
      + \frac{\partial\hfill}{\partial c},
      \quad 
      \InfGen{\delta}_{2} :=
      x\frac{\partial\hfill}{\partial x} - b\frac{\partial\hfill}{\partial b},
      \quad 
      \InfGen{\delta}_{3} :=
      a\frac{\partial\hfill}{\partial a} + b\frac{\partial\hfill}{\partial b}
      + c\frac{\partial\hfill}{\partial c} -t \frac{\partial\hfill}{\partial t}.
    \end{equation}
    The section~\ref{sec:DeterminingSystem} describes the required
    computations and shows that their complexity is quasi-polynomial
    in input size.
  \item[Step~2.]  Choose a generator that is a symmetry of the others
    (above infinitesimal generators form a solvable Lie algebra and
    our reduction process have to take this property into account).
    As~${[\InfGen{\delta}_{1}, \InfGen{\delta}_{2}] =
      [\InfGen{\delta}_{2}, \InfGen{\delta}_{3} ]=0}$
    and~${[\InfGen{\delta}_{1},
      \InfGen{\delta}_{3}]=\InfGen{\delta}_{1}}$, we could
    choose~$\InfGen{\delta}_{1}$ or~$\InfGen{\delta}_{2}$.  Determine
    a \emph{principal element}~(${\principalelement:= \log b}$) of
    this last generator i.e.\ an element defining a coordinates set in
    which the derivation~$\InfGen{\delta}_{2}$ is rectified (equal to
    the translation~${\partial/\partial \principalelement}$).
  \item[Step~3.] The chosen principal element induces an
    \emph{invariantization} of the considered system i.e.\ the
    system~${\dot{\invariant{x}} = (a-\invariant{x})\invariant{x} - c
      \invariant{x}}$ that is invariant under the
    action~${\mathcal{S}_{\mu}: {\freeparam{x} \rightarrow \mu
        \freeparam{x}},\ {\freeparam{b} \rightarrow
        \freeparam{b}/\mu}}$ of the one-parameter group of symmetry
    induced by~$\InfGen{\delta}_{2}$.  The solution~$x$ of the
    original system is then parameterized
    by~$\mathcal{S}_{b}(\invariant{x})$ where~$\invariant{x}$ is a
    solution of its invariantization and~$b$ is a free parameter.
  \item[] Repeat Step~1 (supplementary affine symmetries could
    appear after Step~3).
  \end{itemize}
\end{examplecontinued}
Let us stress that it is generally hard to find a general
infinitesimal generator of a system's symmetry (Step~1) and to give an
explicit representation of an invariant coordinates set (Step~2) for
it.  Thus, we restrict ourself to Lie symmetries associated to 
affine infinitesimal generators for which invariant coordinates
computation is easy (for general case see~\cite{HubertKogan2006} and
references therein). Hence, we do not follows methods developed for
general cases because their complexity are likely exponential in
input's size while we focus our attention to
method of quasi-polynomial complexity.
\par
To conclude, remark that this reduction process works also for purely
algebraic system (describing fixed point of a dynamical system for
example).
\subsubsection{Outline.}
\label{sec:outline}
In the next section, we recall some basic definitions concerning
considered systems and related derivations.  Then, we present the
notion of principal element and show how it could be used in order to
define a rectifying coordinates set for general derivations.  In the
second part of this note, we focus our attention on affine Lie point
symmetries in order to propose a probabilistic strategy to compute
them and their associated principal elements.  We show how previously
introduced notions are used in the reduction process by
considering invariantization of purely algebraic (resp.\ differential)
system and their parameterization.  Finally, in conclusion we make
some remarks and suggest possible further works.
\section{Considered systems and associated derivations}
\label{sec:ConsideredSystemsAndDerivations}
\subsection{Some Algebraic Systems used in Analysis of Biological
  Model}
\label{sec:considered-systems}
\begin{note}
  \label{nota:DifferentialSystem}
  \emph{Notations} --- Hereafter, we consider an explicit algebraic
  ordinary differential system~$\OrdinaryDifferentialSystem$ bearing
  on~$\statevariablenb$ state variables~${\statevariableset := (
    \statevariable_{1}, \dots, \statevariable_{\statevariablenb})}$
  and depending on~$\parameternb$ parameters~${\parameterset := (
    \parameter_{1}, \dots, \parameter_{\parameternb})}$:
  \begin{equation}
    \label{eq:OrdinaryDifferentialSystem} 
    \OrdinaryDifferentialSystem \qquad \left\lbrace
      \begin{array}{l} 
        \dot{\statevariableset} = 
        \VectorFieldSecondMemberset(t,\statevariableset,\parameterset),
        \\
        \dot{\time}  =  1, \quad \dot{\parameterset}  =  0. 
      \end{array} \right.
  \end{equation}
  Denoting the set~${ \{1,\ldots,\statevariablenb\} }$
  by~$\statevariableindiceset$, the letter~$\dot{\statevariableset}$
  stands for first order derivatives of state
  variables~${(\dot{\statevariable}_{\indices}\, |\, \indices \in
    \statevariableindiceset)}$ w.r.t.\ time~$\time$
  and~${\VectorFieldSecondMemberset :=(
    \VectorFieldSecondMember_{\indices} \, |\, \indices \in
    \statevariableindiceset)}$ is a finite subset
  of~$\basefield(\time,\statevariableset,\parameterset)$
  where~$\basefield$ is a subfield ($\field{Q}$ for example)
  of~$\field{C}$.  In order to determine the qualitative properties of
  the dynamical system~(\ref{eq:OrdinaryDifferentialSystem}), it is
  usual to consider the following systems for various
  subset~$\indicesubset$ of~$\statevariableindiceset$:
  \begin{equation}
    \label{eq:InducedOrdinaryDifferentialSystem} 
    \OrdinaryDifferentialSystem_{\indicesubset} \qquad \left\lbrace
      \begin{array}{ll} 
        \dot{\statevariable}_{\indices} = 
        \VectorFieldSecondMember_{\indices}(t,\statevariableset,\parameterset),
        & \quad \forall \indices\in\indicesubset, \\
        \dot{\statevariable}_{\indicep} =  
        \VectorFieldSecondMember_{\indicep}(t,\statevariableset,\parameterset)=0,
        & \quad \forall \indicep \in 
        \statevariableindiceset\setminus\indicesubset, \\
        \dot{\time}  =  1, \quad \dot{\parameterset}  =  0,
      \end{array} \right.
  \end{equation}
  in which some state variables are considered as parameters.  In
  fact, for~$\indicep$ in~$\statevariableindiceset$, the system
  $\OrdinaryDifferentialSystem_{\{\indicep\}}$ defines the
  so-called~$\statevariable_{\indicep}$-nullcline
  of~$\OrdinaryDifferentialSystem$ and the purely algebraic
  system~$\OrdinaryDifferentialSystem_{\emptyset}$ defines its fixed
  points (see examples of applications in~\cite{Murray2002}).
\end{note}
\begin{remark}
  \label{rem:ExpandedStateSpace}
  In the sequel, we are going to avoid---as much as possible---any
  distinction between time, state variables and parameters i.e.\ we
  work in an expanded state space (see~\cite{Burde2002} for another
  application of this standpoint); hence, let us denote the
  set~$(\time,\statevariableset,\parameterset)$
  by~${\variableset:=(\variable_{\indicep}\,|\,1\leq\indicep\leq
    1+\parameternb+\statevariablenb)}$ and its cardinal
  by~$\variablenb$.
\end{remark}
\subsection{Infinitesimal Generators, associated flows and their
  rectification}
\label{sec:Derivation}
First let us recall some basic facts about derivations.
\begin{definition}
  \label{def:derivation}
  Given a polynomial algebra~$\basefield[\variableset]$, a derivation
  of~$\basefield[\variableset]$ with constant field~$\basefield$ is an
  additive mapping~${\InfGen{\delta}: \basefield[\variableset]
    \rightarrow \basefield[\variableset]}$ that satisfies Leibniz
  rules:
  \begin{equation}
    \label{eq:Leibnizrules}
    \forall (\function_{1} , \function_{1}) \in \basefield[\variableset]^2, 
    \quad
    \InfGen{\delta}( \function_{1} \function_{2}) = \
    \function_{1}\InfGen{\delta} \function_{2} 
    + \function_{2} \InfGen{\delta} \function_{1},
  \end{equation}
  and have~$\basefield$ in its kernel.  We denote
  by~${\DerModule{\basefield}{\basefield[\variableset]}}$ the set of
  all such derivations.  The \emph{Lie bracket} is defined by
  the~$\basefield$-bilinear map:
  \begin{equation}
    \label{eq:LieBracketDefinition}
    [\ ,\ ]:
    \begin{array}[t]{ccc}
      \DerModule{\basefield}{\basefield[\variableset]}
      \times 
      \DerModule{\basefield}{\basefield[\variableset]}
      & \to & \DerModule{\basefield}{\basefield[\variableset]}, \\
      (\InfGen{\delta}_{1},\InfGen{\delta}_{2})
      & \to &
      \InfGen{\delta}_{1}\InfGen{\delta}_{2} 
      - \InfGen{\delta}_{2}\InfGen{\delta}_{1}.
    \end{array}
  \end{equation}
  This map is skew-symmetric and satisfies the following Jacobi
  identity:
  \begin{equation}
    \label{eq:JacobiIdentity}
    \forall\ (\InfGen{\delta}_{1},\InfGen{\delta}_{2},
    \InfGen{\delta}_{3}) \subset
    \DerModule{\basefield}{\basefield[\variableset]},
    \quad
    [\InfGen{\delta}_{1}, [\InfGen{\delta}_{2},\InfGen{\delta}_{3}]] 
    + [\InfGen{\delta}_{2}, [\InfGen{\delta}_{3},\InfGen{\delta}_{1}]]
    + [\InfGen{\delta}_{3}, [\InfGen{\delta}_{2},\InfGen{\delta}_{1}]]=0.
  \end{equation}
  The set~${\DerModule{\basefield}{\basefield[\variableset]}}$ is
  a~$\basefield$ vector-space spanned by the set of canonical
  derivations~${\{\partial/\partial
    \variable_{1},\dots,\partial/\partial \variable_{\variablenb}\}}$.
  It is also a Lie algebra with Lie bracket as product.
\end{definition}
\begin{remark}
  An algebraic system~$\OrdinaryDifferentialSystem_{\indicesubset}$
  defined by~(\ref{eq:InducedOrdinaryDifferentialSystem}) could be
  seen as a derivation:
  \begin{equation}
    \label{eq:NullClineInfGen}
    \textstyle
    \InfGen{D}_{\indicesubset} :=
    \frac{\partial\hfill}{\partial\time}+
    \sum_{\indices\in\indicesubset}
    \VectorFieldSecondMember_{\indices}
    \frac{\partial\hfill}{\partial\statevariable_{\indices}},
  \end{equation}
  associated to the algebraic relations~${\{
    \VectorFieldSecondMember_{\indicep} = 0,\, \forall\indicep\in
    \statevariableindiceset\setminus \indicesubset \}}$.
\end{remark}
\subsubsection{Derivations considered as infinitesimal generators.}
The \emph{exponentiation} of a derivation~$\InfGen{\delta}$ induces
several morphisms as shown by following definitions:
\begin{definition}
  \label{def:FormalFlowInfinitesimalGenerator}
  Given a derivation~$\InfGen{\delta}$ and~$\flowtime$ one of its
  constant~(${\InfGen{\delta} \flowtime = 0}$), one can define the
  exponential map~${\ExponentialMap{\flowtime\InfGen{\delta}} :=
    \sum_{\indicep \in \field{N}} \flowtime^{\indicep}
    \InfGen{\delta}^{\indicep}/\indicep!}$
  from~$\basefield[\variableset]$ into the
  algebra~$\basefield[[\flowtime,\variableset]]$ of power series in
  the indeterminates~$(\flowtime,\variableset)$.
  \begin{enumerate}
  \item This map is a morphism that associates to any~$\function$
    in~$\basefield[\variableset]$ its \emph{Lie series} defined by the
    formal power series~${\sum_{\indicep \in \field{N}}
      \flowtime^{\indicep}
      \InfGen{\delta}^{\indicep}\function/\indicep!}$.
  \item The derivation~$\InfGen{\delta}$ is called the
    \emph{infinitesimal generator}
    of~$\ExponentialMap{\flowtime\InfGen{\delta}}$.
  \item The formal power
    series~$\ExponentialMap{\flowtime\InfGen{\delta}} \variableset$
    are solutions of the vector field associated to~$\InfGen{\delta}$.
    These series form the \emph{formal flow} of~$\InfGen{\delta}$;
    this derivation induces an infinitesimal transformation
    from~${\basefield \times \basefield^{\variablenb}}$
    into~$\basefield^{\variablenb}$ that associates, under suitable
    condition of convergence, the
    evaluation~$(\ExponentialMap{\flowtime\InfGen{\delta}}
    \variableset)(\originalsolution)$ to any parameter~$\flowtime$
    in~$\basefield$ and any initial point~$\originalsolution$
    in~$\basefield^{\variablenb}$; this map is the action of the
    flow~$\ExponentialMap{\flowtime\InfGen{\delta}}$
    on~$\basefield^{\variablenb}$.
  \end{enumerate}
\end{definition}
\begin{example}
  \label{ex:InfinitesimalGenerator}
  Hence,
  the~$\field{C}(\statevariable)$-morphism~${\automorphism_{\flowtime}:
    \statevariable \rightarrow e^{\flowtime}\statevariable}$ could be
  defined by the exponential map~${\automorphism_{\flowtime} :=
    \ExponentialMap{\flowtime\InfGen{\delta}}}$ where~$\InfGen{\delta}$
  denotes the derivation~${\statevariable
    \partial/\partial\statevariable}$ acting on the
  field~$\field{C}(\statevariable)$. The
  set~${\{\automorphism_{\flowtime} \,|\, \flowtime \in \field{C}\}}$ is
  a one-parameter group of automorphisms.
\end{example}
\begin{lemma}
  \label{lem:BCH}
  Given two derivations~$\InfGen{\partial}$ and~$\InfGen{\delta}$, the
  Baker Campbell Hausdorff formula states that the relation~${
    \ExponentialMap{\InfGen{\partial}}\ExponentialMap{\InfGen{\delta}}
    = \ExponentialMap{\InfGen{\delta}}
    \ExponentialMap{\InfGen{\partial}}
    \ExponentialMap{[\InfGen{\partial},\InfGen{\delta}]}}$ holds.
\end{lemma}
The next section presents some
how some derivations could be expressed as translation in a suitable
coordinates set.
\subsubsection{Some Algebraic Tools for Rectification of an
  Infinitesimal Generator.}
\label{sec:InfinitesimalGeneratorRectification}
\paragraph{Principal element.} 
Forthcoming manipulations are based on the existence of a special
element that behaves as a \emph{time} variable for considered
derivation as shown by the following definition:
\begin{definition}
  \label{def:PrincipalElement}
  An~$\principalelement$ element in a algebra~$\algebra$ is
  \emph{principal} for a derivation~$\InfGen{\delta}$ acting
  on~$\algebra$ if the relation~${\InfGen{\delta}\principalelement
  =1}$ holds.
\end{definition}
In order to determine a principal element~$\principalelement$ of a
derivation~$\InfGen{\delta}$, one have to solve the following partial
differential equation~${\InfGen{\delta} \principalelement = 1}$. As
this is not a trivial task and as not every derivation has such an
element, we are going in the sequel to restrict our manipulation to
the following kind of principal elements:
\begin{lemma}
  \label{lem:TwoKindOfPrincipalElement}
  Given a derivation~$\InfGen{\delta}$ of~$\basefield[\variableset]$,
  if there exists an element~$\preprincipalelement$
  in~$\basefield[\variableset]$,
  \begin{enumerate}
  \item such that the
    relations~${\InfGen{\delta}\preprincipalelement\not = 0}$
    and~${\InfGen{\delta}^2\preprincipalelement=0}$ hold, then the
    fraction~${\principalelement :=
      \preprincipalelement/\InfGen{\delta}\preprincipalelement}$
  \item and a constant~$\eigenvalue$ of~$\InfGen{\delta}$ such that
    the relation~${\InfGen{\delta} \preprincipalelement = \eigenvalue
      \preprincipalelement}$ holds, then for any constant~$\constant$
    of this derivation, the transcendental element~${\principalelement
      := \log(\constant\preprincipalelement)/\eigenvalue}$
  \end{enumerate}
  is a principal element~$\principalelement$ of~$\InfGen{\delta}$. The
  element~$\preprincipalelement$ is called the \emph{preprincipal
    element} of~$\InfGen{\delta}$ associated to~$\principalelement$.
\end{lemma}
\begin{remark}
  As we adopt an algebraic standpoint in this note, this lemma
  requires to consider in the sequel the
  localization~$\localization{\basefield[\principalelement,
    \variableset]}{\multiplicativeset}$
  of~$\basefield[\principalelement, \variableset]$ at the
  multiplicative closed set~${\multiplicativeset :=
    \{(\InfGen{\delta}\preprincipalelement)^{\indicep}\, |\, \indicep
    \in \integer\}}$ (resp.~${\multiplicativeset :=
    \{\preprincipalelement^{\indicep}\, |\, \indicep \in
    \integer\}}$).  In fact, given any canonical
  derivation~$\partial/\partial\variable$
  of~$\basefield[\variableset]$, there exists one, and only one,
  canonical derivation of~$\localization{\basefield[\principalelement,
    \variableset]}{\multiplicativeset}$
  extending~$\partial/\partial\variable$ and such that the following
  usual relations~${ \lfrac{\partial
      \principalelement}{\partial\variable} = \lfrac{\partial
      \preprincipalelement}{\partial\variable} -
    (\lfrac{\principalelement}{(\InfGen{\delta}\preprincipalelement)^{2}})
    \lfrac{\partial
      \InfGen{\delta}\preprincipalelement}{\partial\variable}}$
  (resp.~${\lfrac{\partial \principalelement}{\partial\variable} =
    (\lfrac{1}{\preprincipalelement}) \lfrac{\partial
      \preprincipalelement}{\partial\variable}}$) are well defined
  in~$\localization{\basefield[\principalelement,
    \variableset]}{\multiplicativeset}$ (for the sake of simplicity,
  we use the same notation for derivations acting
  on~$\basefield[\variableset]$ and their extension to derivation
  acting on~$\localization{\basefield[\principalelement,
    \variableset]}{\multiplicativeset}$).  This shows that the Lie
  algebra~${\DerModule{\basefield}
    {\localization{\basefield[\principalelement,\variableset]}
      {\multiplicativeset}}}$ is well defined.
\end{remark}
\begin{example} For any
  element~$\fractionfieldelement$ in~$\basefield(\variableset)$, we
  consider the logarithm~$\log\fractionfieldelement$ i.e\ a
  transcendental field
  extension~$\basefield(\variableset,\log\fractionfieldelement)$ and
  the associated derivation extension such
  that~${\InfGen{\delta}\log\fractionfieldelement =
    \InfGen{\delta}\fractionfieldelement/\fractionfieldelement}$.
  Hence, the derivation~${\InfGen{\delta} :=
    \statevariable\partial/\partial\statevariable}$ acting
  on~$\field{C}(x)$ has a unique extension to a
  derivation~$\overline{\InfGen{\delta}}$ acting
  on~$\field{C}(x,\log(x))$ such that the
  relation~${\overline{{\InfGen{\delta}}}\log(x)=1}$ holds.
\end{example}
\paragraph{Construction of a Rectifying Coordinate Ring.}
\label{sec:ConstructionOfRectifyingCoordinateRing}
Principal elements of a derivation~$\InfGen{\delta}$ allow to
construct a rectifying field in which~$\InfGen{\delta}$ acts as a
simple translation.
\begin{lemma}
  \label{lem:InducedProjection}
  Given a derivation~$\InfGen{\delta}$ and one of its principal
  element~$\principalelement$, let us define the following formal
  operator:
  \begin{equation}
    \label{eq:RectifyingOperator}
    \projection{\InfGen{\delta}}{\principalelement} := 
    \sum_{\indicep\in\mathbb{N}}
    (-\principalelement)^{\indicep}
    \frac{\InfGen{\delta}^{\indicep}}{\indicep!}.
  \end{equation}
  As~$\InfGen{\delta}$ is a derivation, this operator induces a
  homomorphism and the following exact sequence:
  \begin{equation}
    \label{eq:ExactSequence}
    0 \rightarrow
    \ker \projection{\InfGen{\delta}}{\principalelement}
    \rightarrow
    \localization{\basefield[\principalelement,\variableset]}
    {\multiplicativeset}
    \stackrel{\projection{\InfGen{\delta}}{\principalelement}}
    {\longrightarrow}
    \localization{\basefield[[\principalelement,\variableset]]}
    {\multiplicativeset}
    \rightarrow
    \basefield[\invariantset]
    \rightarrow
    0
  \end{equation}
  where the variables set~$\invariantset$ denotes the
  set~$\projection{\InfGen{\delta}}{\principalelement}\variableset$ of
  formal power series.
\end{lemma}
\begin{remark}
  \label{rem:ConstantSubFieldOfRectifyingField}
  To prove that the
  map~$\projection{\InfGen{\delta}}{\principalelement}$ is a
  homomorphism, one can use the same argument then whose used in the
  proof stating the same property for the exponential
  map~$\ExponentialMap{\InfGen{\delta}}$.  By
  construction~$\projection{\InfGen{\delta}}{\principalelement}
  \principalelement$ is equal to~$0$. Thus, the kernel~$\ker
  \projection{\InfGen{\delta}}{\principalelement}$ contains the
  ideal~$\principalelement \localization
  {\basefield[\principalelement,\variableset]} {\multiplicativeset}$
  and is not trivial (see also Proposition~\ref{prop:CoordinatesChange}).
  \par
  Using exact sequence~(\ref{eq:ExactSequence}), we could define a
  coordinate ring~$\basefield[\invariantset]$ that is isomorphic to
  the quotient algebra~${
    \localization{\basefield[\principalelement,\variableset]}
    {\multiplicativeset} / ( \ker
    \projection{\InfGen{\delta}}{\principalelement})
    \localization{\basefield[\principalelement,\variableset]}
    {\multiplicativeset}}$ and a \emph{rectifying}
  ring~$\basefield[\invariantset,\principalelement]$ that is its
  finitely generated extension.  To explain this terminology, first
  remark that the derivation~$\InfGen{\delta}$ acting
  on~$\localization{\basefield[\principalelement,\variableset]}
  {\multiplicativeset}$ could be easily extended to a derivation
  acting on~$
  \localization{\basefield[[\principalelement,\variableset]]}
  {\multiplicativeset}$ and thus
  to~$\basefield[\principalelement,\invariantset]$.  The following
  lemma states that the derivation~$\InfGen{\delta}$ is
  \emph{rectified} when we consider its action
  on~$\basefield[\principalelement, \invariantset]$:
  \begin{lemma}
    \label{lem:DeltaRectification}
    With previously introduced notations, the following relations
    hold:
    \begin{equation}
      \label{eq:Rectification}
      \InfGen{\delta}\principalelement = 1, 
      \quad
      \InfGen{\delta}\invariantset = 0.
    \end{equation}
  \end{lemma}
  \begin{sketchofproof}
    The first relation is the definition of a principal element.
    Elements~$\invariantset$ are defined by the
    series~${\sum_{\indicep \in \mathbb{N}}
      (-\principalelement)^{\indicep} \InfGen{\delta}^{\indicep}
      \variableset/\indicep!}$. By Leibniz' rule, we have:
    \begin{equation}
      \label{eq:PrincipalElementTrick}    
      \textstyle
      \forall \indicep \in \mathbb{N},\quad 
      \InfGen{\delta}\left( (-\principalelement)^{\indicep}
        \frac{\InfGen{\delta}^{\indicep}}{\indicep!} 
      \right)= (-1)^i\left(
        \principalelement^{\indicep}
        \frac{\InfGen{\delta}^{\indicep+1}}{\indicep!} +
        \principalelement^{\indicep-1}
        \frac{\InfGen{\delta}^{\indicep}}{(\indicep-1)!}
      \right)\!,
    \end{equation}
    and thus, derivation's linearity proves the last relations.
  \end{sketchofproof}
  The morphism~$\projection{\InfGen{\delta}}{\principalelement}$
  induces a coordinates change allowing to express the
  derivation~$\InfGen{\delta}$ as a simple
  translation~$\partial/\partial \principalelement$ in this new
  coordinates set.  We do not describe further this coordinates
  change, because we are just going to use some of its properties and
  not its exact formulation.  Forthcoming considerations are based on
  the fact that the relations~${\InfGen{\delta}\invariantset = 0}$
  imply that the relations~${\ExponentialMap{\flowtime\InfGen{\delta}}
    \invariantset = \invariantset}$ hold. Thus, the
  morphism~$\projection{\InfGen{\delta}}{\principalelement}$ maps the
  coordinate ring~$\basefield[\variableset]$ of the ambient
  space~$\basefield^{\variablenb}$ onto a coordinate ring invariant
  under the action of the
  flow~$\ExponentialMap{\flowtime\InfGen{\delta}}$.
\end{remark}
\subsection{Expanded Lie Point Symmetries and their Determining
  System}
\label{sec:InfGenAndDetSyst}
Let us define now the derivations used in the sequel.
\begin{definition}
  \label{def:LiePointSymmetry}
  Given a derivation~$\InfGen{\delta}$, an algebraic
  system~$\OrdinaryDifferentialSystem_{\indicesubset}$ and the
  associated derivation~$\InfGen{D}_{\indicesubset}$,
  $\InfGen{\delta}$ is an infinitesimal generator of an \emph{expanded
    Lie point symmetry}
  of~$\OrdinaryDifferentialSystem_{\indicesubset}$ if there exists a
  constant~$\lambda$ of~$\InfGen{D}_{\indicesubset}$ such that the
  following relations hold:
  \begin{eqnarray}
    \label{eq:GeneralLieDetermingSystemTime}
    \InfGen{D}_{\indicesubset}\InfGen{\delta}(\time) = 
    \frac{\partial \InfGen{\delta}(\time)}{\partial \time} 
    +\sum_{\indices\in\indicesubset}\VectorFieldSecondMember_{\indices}
    \frac{\partial \InfGen{\delta}(\time)}{\partial \statevariable_{\indices}}
    &=& -\lambda,\\
    \label{eq:GeneralLieDetermingSystemStateVariables}
    \sum_{\variable\in\variableset}\InfGen{\delta}(\variable)
    \frac{\partial \VectorFieldSecondMember_{\indicep}}{\partial \variable}
    -\frac{\partial \InfGen{\delta}(\statevariable_{\indicep})}{\partial \time}
    -\sum_{\indices\in\indicesubset} \VectorFieldSecondMember_{\indices}
    \frac{\partial \InfGen{\delta}(\statevariable_{\indicep})}
    {\partial \statevariable_{\indices}}
    &=&\lambda \VectorFieldSecondMember_{\indicep}, 
    \quad \forall \indicep\in \indicesubset,\\  
    \label{eq:GeneralLieDetermingSystemPurelyAlgebraicEquation}
    \InfGen{\delta}\VectorFieldSecondMember_{\indicep} =
    \sum_{\variable \in\variableset}
    \InfGen{\delta}(\variable)\frac{\partial \VectorFieldSecondMember_{\indicep}}
    {\partial\variable} &=& 
    \lambda \VectorFieldSecondMember_{\indicep}, 
    \quad \forall \indicep\in\statevariableindiceset\setminus\indicesubset, 
    \\
    \label{eq:GeneralLieDetermingSystemParameters}
    \InfGen{D}_{\indicesubset}\InfGen{\delta}(\parameter) =
    \frac{\partial\InfGen{\delta}(\parameter)}{\partial \time}
    + \sum_{\indices\in\indicesubset}\VectorFieldSecondMember_{\indices}
    \frac{\partial\InfGen{\delta}(\parameter)}{\partial \statevariable_{\indices}}
    &=&0,\quad  \forall \parameter\in \parameterset.
  \end{eqnarray}
  These relations form the \emph{determining system}
  of~$\OrdinaryDifferentialSystem_{\indicesubset}$ expanded Lie point
  symmetries.
\end{definition}
\begin{remark}
  \label{rem:LieAlgebraChain}
  \emph{Solution space structure} --- Derivations~$\InfGen{\delta}$
  satisfying~(\ref{eq:GeneralLieDetermingSystemTime})\,--%
  \,(\ref{eq:GeneralLieDetermingSystemParameters}) form a Lie
  sub-algebra of~$\DerModule{\field{A}}{\basefield[\variableset]}$
  denoted by~$\Liealgebra{\OrdinaryDifferentialSystem}$.  Furthermore,
  if~$\indicesubset_{1}$ is a subset of~$\indicesubset_{2}$, the Lie
  algebra~$\Liealgebra{\OrdinaryDifferentialSystem_{\indicesubset_{1}}}$
  is a sub-algebra
  of~$\Liealgebra{\OrdinaryDifferentialSystem_{\indicesubset_{2}}}$.
\end{remark}
\begin{remark}
  \label{rem:GeneralDeterminingSystemsAreUnsolvable}
  \emph{Considered Lie symmetries vs general Lie symmetries} --- The
  definition~\ref{def:LiePointSymmetry} is designed for our algebraic
  purposes but is only a restriction of the general definition of Lie
  point symmetries (see~\cite{Olver1993}). In fact, remark that if the
  considered algebraic system~$\OrdinaryDifferentialSystem_{\indicesubset}$ is
  \begin{itemize}
  \item a vector field~(${\indicesubset=\statevariableindiceset}$),
    this definition reduces to the classical one of Lie point
    symmetries based on the Lie bracket
    i.e.~${[\InfGen{D},\InfGen{\delta}]= \lambda\InfGen{D}}$;
  \item a purely algebraic system~(${\indicesubset=\emptyset}$), this
    definition is more restrictive than the classical definition
    presented in Section~2.1 of~\cite{Olver1993}. In fact, let us
    consider the following system~${ f_{1} :=
    {x_{1}}^{\!2}+{y_{1}}^{\!2}-1,\ f_{2} :=
    {x_{2}}^{\!2}+{y_{2}}^{\!2}-1,\ f_{3} := x_{2}y_{1} -
    y_{2}x_{1}}$ and the derivation~${ \InfGen{\delta} :=
    x_{2}\lfrac{\partial\hfill}{\partial y_{1}} -
    y_{2}\lfrac{\partial\hfill}{\partial x_{1}} +
    x_{1}\lfrac{\partial\hfill}{\partial y_{2}} -
    y_{1}\lfrac{\partial\hfill}{\partial x_{2}}}$.  
    The relations~${\InfGen{\delta}f_{1}=2f_{3},\
      \InfGen{\delta}f_{2}= -2f_{3},\
      \InfGen{\delta}f_{3}=f_{1}-f_{2}}$ hold i.e.\  the
    derivation~$\InfGen{\delta}$ leaves invariant the ideal spanned
    by~$f_{1},f_{2},f_{3}$ and thus according
    to~\cite{Olver1993},~$\InfGen{\delta}$ is the infinitesimal
    generator of a one-parameter group of Lie symmetry (a family of
    morphism~$\ExponentialMap{\flowtime\InfGen{\delta}}$ parameterized
    by a constant~$\flowtime$) that leaves the ideal spanned
    by~$\{f_{1}, f_{2}, f_{3}\}$ invariant but not each of these
    polynomials (as shown by the relation~${ e^{\flowtime
        \InfGen{\delta}} f_{1} = f_{1} \cos^{2} \flowtime + f_{2}
      \sin^{2} \flowtime + 2 f_{3} \cos \flowtime \sin \flowtime}$).
    This kind of derivations are not taken into account in this note.  This restriction is motivated by
    computational purposes i.e.\ the general definition of a Lie point
    symmetry for an algebraic system~${F=0}$ implies that we
    could---at least---determine~$0$ in the quotient
    algebra~$\basefield[\variableset]/F\basefield[\variableset]$; as
    this task is not in the complexity class considered in this note,
    we made a first restriction to the set of Lie symmetry used in our
    work by only considering solution of
    system~(\ref{eq:GeneralLieDetermingSystemTime})\,--%
    \,(\ref{eq:GeneralLieDetermingSystemParameters}).
  \end{itemize}  
\end{remark}
  Furthermore, there is little hope to solve the general partial
  differential problem~(\ref{eq:GeneralLieDetermingSystemTime})\,--%
  \,(\ref{eq:GeneralLieDetermingSystemParameters}); thus, we restrict
  our solution space to \emph{affine} infinitesimal generators for
  which the associated determining equations form a linear system.
\section{Affine derivation and associated invariantization}
\label{sec:affine-infin-gener}
\begin{definition}
  Let us denote by~$\AffDer{\basefield}{\basefield[\variableset]}$ the
  following set of derivations:
  \begin{equation}
    \label{eq:DerivationsModule}
    \left\lbrace 
      \InfGen{\delta} =\!  \sum_{\variable\in\variableset} \InfGen{\delta}(\variable)
      \frac{\partial\hfill}{\partial \variable}
      \, \Big |\,
      \InfGen{\delta}(\variable_{1}) := b_{\variable_{1}} 
      \!\!+\!\!
      \sum_{\variable_{2} \in \variableset}
      \!\!a_{\variable_{1}\variable_{2}}\variable_{2},\,
      \big(
      b_{\variable_{1}}, a_{\variable_{1}\variable_{2}}\,|\, \variable_{2}\in\variableset
      \big)\in
      \basefield^{\variablenb+1} 
      \!\right\rbrace\!.
  \end{equation}
\end{definition}
\begin{note}
  \label{nota:MatrixNotation}
  \emph{Notation} --- Given a derivation~$\InfGen{\delta}$
  in~$\AffDer{\basefield}{\basefield[\variableset]}$, we are going in
  the sequel to consider~$\variableset$ as a vector and use the
  following matricial notations:
  \begin{equation}
    \label{eq:MatricesNotations}
    \LinearPart{\InfGen{\delta}} = 
    \left( 
      a_{\variable_{1}\variable_{2}}
    \right)_{({\variable_{1},\variable_{2}})\in\variableset^{2}},
    \qquad
    \TransPart{\InfGen{\delta}} 
    = \left( b_{\variable} \right)_{\variable\in\variableset},
    \qquad
    \InfGen{\delta}\variableset=
    \LinearPart{\InfGen{\delta}} \variableset +
    \TransPart{\InfGen{\delta}}.
  \end{equation}
\end{note}
\subsection{Determining System defining Affine Infinitesimal generators}
\label{sec:DeterminingSystem}
\begin{lemma}
  For an affine infinitesimal generator~$\InfGen{\delta}$
  in~$\AffDer{\basefield}{\basefield[\variableset]}$, the associated
  determining system~(\ref{eq:GeneralLieDetermingSystemTime})\,--%
  \,(\ref{eq:GeneralLieDetermingSystemParameters}) reduces to the
  following linear system:
  \begin{equation}
    \label{eq:LinearDeterminingSystem}
    \left(
      \begin{array}{ccc}
        0 & \cdots & 0 \\
        \frac{\partial f_{1}}{\partial \variable_{1}} & \cdots &
        \frac{\partial f_{1}}{\partial \variable_{\variablenb}} \\
        \vdots & & \vdots \\
        \frac{\partial f_{\statevariablenb}}{\partial \variable_{\variablenb}} &
        \cdots &
        \frac{\partial f_{\statevariablenb}}{\partial \variable_{\variablenb}} \\
        0 & \cdots & 0 \\
        \vdots & & \vdots \\
        0 & \cdots & 0 
      \end{array}
    \right)
    ( \LinearPart{\InfGen{\delta}} \variableset + \TransPart{\InfGen{\delta}})
    - \LinearPart{\InfGen{\delta}}
    \left(
      \begin{array}{c}
        1 \\
        f_{1} \caracteristic{1\in\indicesubset}\\
        \vdots \\
        f_{\statevariablenb}\caracteristic{\statevariablenb\in\indicesubset} \\
        0 \\
        \vdots \\
        0 
      \end{array}
    \right) =
    \lambda \left(
      \begin{array}{c}
        1 \\
        f_{1} \\
        \vdots \\
        f_{\statevariablenb} \\
        0 \\
        \vdots \\
        0 
      \end{array}
    \right)\!,
  \end{equation}
  where~${\caracteristic{\indicep\in\indicesubset}}$ is equal to~$1$
  if the index~$\indicep$ is in~$\indicesubset$ and~$0$ otherwise.
\end{lemma}
\begin{remark}
  \label{rem:ShortVectorChoice}
  \emph{Probabilistic resolution of determining system defining affine
    derivation} --- The system~(\ref{eq:LinearDeterminingSystem})
    could be rewritten in a the more convenient matricial
    notation~${M(\variableset)K=0}$ where~$M(\variableset)$ is
    a~${\variablenb\times(\variablenb+1)\variablenb}$ matrix with
    coefficients in~$\basefield[\variableset]$ and~$K$ is a vector
    whose~${(\variablenb+1)\variablenb}$ coefficients are the
    coefficients of~$\LinearPart{\InfGen{\delta}}$
    and~$\TransPart{\InfGen{\delta}}$.  Affine derivations that are 
    solution of this determining system, are given by the kernel
    of~$M(\variableset)$ in a field~$\basefield$.  Kernel computation
    could be done by the following probabilistic method.
    Indeterminates~$\variableset$ are specialized in
    matrix~$M(\variableset)$ to some random value
    in~$\basefield^{\variablenb}$ in order to obtain a matrix~$M_{1}$
    over the field~$\basefield$; the resulting linear system~$M_{1}K$
    could be underdetermined and thus, several specializations should
    be considered in order to obtain a linear system~$L_{\indicep}$
    defined by~${M_{1}K=\cdots=M_{\indicep}K=0}$. The
    rank~$r_{\indicep}$ of~$L_{\indicep}$ increases with~$\indicep$
    and the specialization process could be stopped
    when~${r_{\indices}=r_{\indices+1}}$; the considered
    system~$L_{\indices}$ could then be solved using a numerical
    method. The specialization set for which this process fails to find
    a correct solution is a zero-dimensional algebraic variety and
    thus, its probability of failure is low.  \par However, there is
    an infinite way to choose a basis of the kernel computed
    above. But, one can use Lenstra, Lenstra and Lov\'asz' basis
    reduction algorithm in order to obtain a reduced basis in the
    sense that less variables are involved in each infinitesimal
    generators definition.  \par To conclude, remark that some
    solutions of system~(\ref{eq:LinearDeterminingSystem}) are
    spurious for our purposes since they describe the same flow and
    should be discarded.  In fact, consider the
    problem~$\OrdinaryDifferentialSystem$ defined
    by~${\dot{\statevariable} = \parameter\statevariable}$: the base
    field of~$\Liealgebra{\OrdinaryDifferentialSystem}$ is the
    constant field of the derivation~${\InfGen{D}:=\partial/\partial
    \time + \parameter\statevariable\partial/\partial\statevariable}$;
    thus, as~$\partial/\partial t$ is
    in~$\Liealgebra{\OrdinaryDifferentialSystem}$ and~$\parameter$ is
    a constant of~$\InfGen{D}$,~$\parameter\partial/\partial t$ is
    another infinitesimal generators representing the same Lie
    symmetry then~$\partial/\partial t$.  These two derivations define
    the same Lie symmetry but are given by two different solutions of
    system~(\ref{eq:LinearDeterminingSystem}).
\end{remark}
\subsection{Principal Element Computation for Affine Derivation}
\label{sec:PrincipalElementComputation}
\begin{lemma}
  \label{lem:PrincipalElementOfAffineDerivation}
  Given a derivation~$\InfGen{\delta}$
  in~$\AffDer{\basefield}{\basefield[\variableset]}$. If there exits a
  vector of~$\InfGen{\delta}$'s constants denoted
  by~${\constantvector:=(\constant_{1}, \ldots,
    \constant_{\variablenb})}$ such that the relations
  \begin{enumerate}
  \item~${{}^{t}\constantvector \LinearPart{\InfGen{\delta}}=0}$
    and~${{{}^{t}\constantvector \TransPart{\InfGen{\delta}}}\not =
      0}$ hold, then the
    fraction~${{}^{t}\constantvector\variableset/{}^{t}\constantvector
      (\LinearPart{\InfGen{\delta}} \variableset +
      \TransPart{\InfGen{\delta}})}$
  \item~${{}^{t}}\LinearPart{\InfGen{\delta}} \constantvector =
    \eigenvalue \constantvector$ and~${\InfGen{\delta}\eigenvalue=0}$
    hold, then the element~${(\log {}^{t}\constantvector (\variableset
      + \TransPart{\InfGen{\delta}}/\eigenvalue))/\eigenvalue}$
  \end{enumerate}
  is a principal element of~$\InfGen{\delta}$.
\end{lemma}
\begin{sketchofproof}
  1) Consider the polynomial~${}^{t}\constantvector\variableset$
  denoted by~$\preprincipalelement$.  Using
  notation~(\ref{eq:MatricesNotations}), remark
  that~$\InfGen{\delta}\preprincipalelement$ is equal to the linear
  combination~${ {}^{t}\constantvector( \LinearPart{\InfGen{\delta}}
  \variableset + \TransPart{\InfGen{\delta}}) }$.  Thus, the
  conditions on~$\constantvector$ given in the first item show
  that~$\InfGen{\delta}\preprincipalelement$ is a constant different
  of~$0$ and thus~$\delta^{2}\preprincipalelement$ is equal to~$0$.
  The first assertion of Lemma~\ref{lem:TwoKindOfPrincipalElement} is
  sufficient to conclude in that case.  2) Consider the
  polynomial~${}^{t}\constantvector (\variableset +
  \TransPart{\InfGen{\delta}}/\eigenvalue)$ denoted
  by~$\preprincipalelement$.  With the hypothesis of the second item,
  the element~$\InfGen{\delta}\preprincipalelement$ is equal
  to~$\eigenvalue\preprincipalelement$ and thus, the second assertion
  of lemma~\ref{lem:TwoKindOfPrincipalElement} is satisfied and the
  transcendental element~${\log(\preprincipalelement)/\eigenvalue}$ is
  a principal element of~$\InfGen{\delta}$.
\end{sketchofproof}
\begin{remark}
  \emph{Computational strategy} --- This lemma shows that in order to
  find a principal element for a derivation~$\InfGen{\delta}$
  in~$\AffDer{\basefield}{\basefield[\variableset]}$, one have first
  to check condition~1) and if it is not satisfied, one have to find
  an eigenvector of~$\LinearPart{\InfGen{\delta}}$.
\end{remark}
\subsection{Flow of a Affine Derivation and Resulting Quotient Space}
\label{sec:ProjectionComputation}
Finding a coordinates change required to place a given derivation in
rectified form is essentially the same problem as solving it in the
first place. This could be easily done for affine derivation using the
Jordan normal form as shown bellow.
\begin{remark}
  \label{rem:JordanNormalForm}
  \emph{Jordan normal form} --- Given a~${\variablenb \times
    \variablenb}$-matrix~$\LinearPart{\delta}$ associated to a
  derivation~$\InfGen{\delta}$, if its minimal polynomial~$p(\xi)$
  is~$\prod_{\indicep=1}^{w} p_{\indicep}$
  with~${p_{\indicep}=(\xi-\eigenvalue_\indicep)^{\eigenmultiplicity_{\indicep}}}$
  and~${\sum_{\indicep=1}^{w}\eigenmultiplicity_{\indicep} =
    \variablenb}$, then there exists a change of
  coordinates~$\VariableChangeMatrix$ such that:
  \begin{equation}
    \label{eq:JordanNormalMatrixForm}
    \textstyle
    \LinearPart{\delta}=\VariableChangeMatrix
    \left(
      \begin{array}{cccc}
        J_{1}   & 0     & \cdots &     0  \\
        0      & J_{2}  & \ddots & \vdots  \\
        \vdots & \ddots & \ddots & 0      \\ 
        0      & \cdots &   0    & J_{w}
      \end{array}
    \right)
    \VariableChangeMatrix^{-1},\
    \textup{with}\ J_{\indicep} := \eigenvalue_{\indicep}
    \textup{Id}_{\eigenmultiplicity_{\indicep}\times\eigenmultiplicity_{\indicep}}+
    \left(
      \begin{array}{ccccc}
        0      & 1  & 0                     & \cdots & 0      \\
        0      & 0  & 1                     & \ddots & \vdots \\
        \vdots &    & \ddots                & \ddots & 0      \\ 
        0      & \multicolumn{2}{c}{\cdots} & 0      & 1      \\ 
        0      & \multicolumn{2}{c}{\cdots} & 0      & 0
      \end{array}
    \right)
  \end{equation}
  if~$\eigenvalue_{\indicep}$ is different from~$0$
  and~${J_{\indicep}:=\eigenvalue_{\indicep}
    \textup{Id}_{\eigenmultiplicity_{\indicep}\times\eigenmultiplicity_{\indicep}}}$
  otherwise; the symbol~$\textup{Id}_{\eigenmultiplicity_{\indicep}
    \times \eigenmultiplicity_{\indicep}}$ denotes the
  identity~$\eigenmultiplicity_{\indicep} \times
  \eigenmultiplicity_{\indicep}$-matrix.  This canonical form and
  Lemma~\ref{lem:PrincipalElementOfAffineDerivation} allow to compute
  preprincipal and associated principal element for any affine
  derivation~$\InfGen{\delta}$.
\end{remark}
\begin{hypotheses}
  From now, we suppose that the base field~$\basefield$
  is~$\mathbb{C}$ in order to contain all eigenvalues of the
  matrix~$\LinearPart{\InfGen{\delta}}$ and to define the quantities
  related to principal elements (exponentials and logarithms).
\end{hypotheses}
Given an affine derivation~$\InfGen{\delta}$, one of its preprincipal
element~$\preprincipalelement$ and the associated principal
element~$\principalelement$, let us interpret geometrically the
manipulation done in
section~\ref{sec:InfinitesimalGeneratorRectification}.
\subsubsection{Flow associated to a derivation and induced equivalence
  classes.}
\label{rem:QuotientByOrbits}
To do so and following
Definition~\ref{def:FormalFlowInfinitesimalGenerator}-3, we consider
the application defined by the linear system of ordinary differential
equations associated to our affine derivation~$\InfGen{\delta}$:
\begin{equation}
  \label{eq:InverseAffineRectifyingOperator}
  \Psi : 
  \begin{array}[t]{cccc}
    {\basefield \times \basefield^{\variablenb}} & \rightarrow 
    & \basefield^{\variablenb}, \\
    (\flowtime,\inducedsolution) & \rightarrow 
    &
    (e^{\flowtime\InfGen{\delta}}\variableset)(\inducedsolution) 
    &
    =
    \exp({ \flowtime \LinearPart{\InfGen{\delta}} }) 
    \inducedsolution + \int_{0}^{\flowtime}
    \exp\bigl({ (\flowtime -\freeparam{s}) \LinearPart{\InfGen{\delta}} }
    \bigr)  
    \TransPart{\InfGen{\delta}} 
    \textup{d}\freeparam{s}.
  \end{array}
\end{equation}
this application could be computed numerically or using the following
relations:
\begin{equation}
  \label{eq:ExpJordanNormalMatrixForm}
  \textstyle
  \exp(\flowtime J_{\indicep}) =        
  \exp(\flowtime\eigenvalue_{\indicep}) \left(\!
    \begin{array}{ccccc}
      1 & 
      \flowtime & \frac{\principalelement^2}{2} &  \cdots & 
      \frac{\flowtime^{\eigenmultiplicity_{i}-1}}
      {(\eigenmultiplicity_{i}-1)!}\\
      0 & 1 & 
      \flowtime & \ddots & \vdots  \\
      \vdots&  & \ddots & \ddots & \frac{\flowtime^2}{2}   \\ 
      \vdots&  & & \ddots &  \flowtime  \\ 
      0     & \multicolumn{2}{c}{\cdots}&0&  
      1      \end{array}
    \!\right)\!.
\end{equation}
\begin{note}\emph{Orbits of the flow --- }
  The image of~${\basefield \times \inducedsolution}$ by~$\Psi$
  constitutes an orbit of the flow~$\ExponentialMap{\flowtime
  \InfGen{\delta}}$.  This standpoint induces an equivalence
  relation~$\sim$ among the point of~$\basefield^{\variablenb}$,
  with~$\originalsolution_{1}$ being equivalent
  to~$\originalsolution_{2}$ if these points lie in the same orbit
  of~$\Psi$.  Let us denotes by~$\hyperplan$ the set of equivalence
  classes a.k.a.\ the set of orbits.  In the sequel, we suppose for
  the sake of conciseness that all the orbits have the same dimension
  i.e.\ we implicitly exclude from our statements the
  lower-dimensional orbits associated to the variety
  in~$\basefield^{\variablenb}$ defined by the ideal~${\{
  \InfGen{\delta}(\variable) = 0 \, |\, \variable \in \variableset \}}$.
\end{note}
All forthcoming manipulations rely on the following remark:
\begin{remark}
  \emph{Invariantization --- } Any object (algebraic relations,
  derivations, etc.) that is invariant under the action of the
  flow~$\ExponentialMap{\flowtime \InfGen{\delta}}$ will have a
  counterpart on the lower-dimensional variety~$\hyperplan$ whose
  representation---the \emph{invariantization} of the considered
  object---completely characterize the original object.
\end{remark}
\begin{note}
  As a first illustration, let us remark that any
  function~${\function : \basefield^{\variablenb} \rightarrow
  \basefield}$ invariant for the flow~$\ExponentialMap{\flowtime
  \InfGen{\delta}}$ is invariant along its orbits and therefore there
  is a well-defined induced function~${\tilde{\function}:
  \hyperplan \rightarrow \basefield}$; conversely, given a
  function~${\tilde{\function}: \hyperplan \rightarrow
  \basefield}$, there is an invariant
  function~${\function: \basefield^{\variablenb} \rightarrow
  \basefield}$ defined by the
  relation~${\function(\originalsolution) := \tilde{\function}(h)}$
  if~$\originalsolution$ is in the orbit~$h$. Hence, we obtain the
  following result:
  \begin{lemma}
    \label{lem:QuotientByOrbits}
    There is a one-to-one correspondence between (polynomial)
    functions on~$\basefield^{\variablenb}$ invariant under the action
    of the flow~$\ExponentialMap{\flowtime \InfGen{\delta}}$ and
    arbitrary (polynomial) functions on~$\hyperplan$.
  \end{lemma}
\end{note}
In order to represent~$\hyperplan$, one can first find an algebraic
representation of the orbits of the flow and then use a cross section
of these orbits (see~\cite{HubertKogan2006} and references therein for
more details).  This general approach is based on Gr\"obner basis
computation and treats general problems that exceed the scope of this
note. Instead, we are going to use Lemmas~\ref{lem:InducedProjection}
and~\ref{lem:QuotientByOrbits} in order to give an algebraic
description of~$\hyperplan$ whose computation is based mainly on
Jordan decomposition.
\subsubsection{Algebraic Representation of~$\hyperplan$.}
Consider the formal
operator~$\projection{\InfGen{\delta}}{\principalelement}$ introduced
in Lemma~\ref{lem:InducedProjection}.
Lemma~\ref{lem:DeltaRectification} implies that the image
of~$\projection{\InfGen{\delta}}{\principalelement}$ is invariant
under the action the flow~$\ExponentialMap{\flowtime
  \InfGen{\delta}}$.  Thus, by describing the kernel
of~$\projection{\InfGen{\delta}}{\principalelement}$, we obtain an
algebraic description of functions on~$\basefield^{\variablenb}$
invariant under the action of the flow~$\ExponentialMap{\flowtime
  \InfGen{\delta}}$.  Lemma~\ref{lem:QuotientByOrbits} shows that this
description induces an algebraic representation of~$\hyperplan$.  The
following lemma recapitulates these points when~$\InfGen{\delta}$ is
an affine derivation:
\begin{proposition}
  \label{prop:CoordinatesChange}
  Given an affine derivation~$\InfGen{\delta}$, one of its
  preprincipal element~$\preprincipalelement$ and the associated
  principal element~$\principalelement$, one can define a
  homomorphism~$\projection{\InfGen{\delta}}{\principalelement}$ on an
  algebra of constants~$\basefield[\invariantset]$
  of~$\InfGen{\delta}$ using the formal
  operator~(\ref{eq:RectifyingOperator}) as follow:
  \begin{equation}
    \label{eq:AffineExactSequence}
    \begin{array}{ccccc}
      0 \rightarrow
      \slice_{\principalelement}
      \localization{\basefield[\variableset]}
      {\multiplicativeset}
      \rightarrow & 
      \localization{\basefield[\variableset]}
      {\multiplicativeset} &
      \stackrel{\projection{\InfGen{\delta}}{\principalelement}}
      {\longrightarrow} &
      \basefield[\invariantset] &
      \rightarrow
      0 \\
      & \polynomialfunction(\variableset) &
      \longrightarrow &
      \polynomialfunction(\invariantset)
    \end{array}
  \end{equation}
  where~$\invariantset$ is equal
  to~$\projection{\InfGen{\delta}}{\principalelement}\variableset$
  and~$\slice_{\principalelement}$ is equal
  to~${}^{t}\constantvector\variableset$ if the preprincipal
  element~$\preprincipalelement$ is defined by the case~1) in
  Lemma~\ref{lem:PrincipalElementOfAffineDerivation} and
  to~${{}^{t}\constantvector (\variableset +
    \TransPart{\InfGen{\delta}}/\eigenvalue)-1}$ otherwise.  The set
  of equivalence classes~$\hyperplan$ could be identify with the
  hyperplane~$V(\slice_{\principalelement})$ of
  dimension~${\variablenb-1}$ defined in~$\basefield^{\variablenb}$ by
  the linear form~$\slice_{\principalelement}$. Furthermore,
  \begin{enumerate}
  \item the map~$\projection{\InfGen{\delta}}{\principalelement}$
    induces a projection that associates to any
    point~$\originalsolution$ in~$\basefield^{\variablenb}$
    s.t.~${\multiplicativeset(\originalsolution)\not = 0}$, the
    point~${ (\projection{\InfGen{\delta}}{\principalelement}
      \variableset )(\originalsolution) :=
      \sum_{\indicep\in\mathbb{N}} \left(
        (-\principalelement(\variableset))^{\indicep}
        \lfrac{(\InfGen{\delta}^{\indicep}Z)}{\indicep!}
      \right)\!(\originalsolution)}$
    of~$V(\slice_{\principalelement})$.
  \item the points composing an orbit
    of~$\ExponentialMap{\flowtime\InfGen{\delta}}$ are projected to a
    single point in~$V(\slice_{\principalelement})$.
  \item the orbit of~$\ExponentialMap{\flowtime\InfGen{\delta}}$
    passing through a point in~$V(\slice_{\principalelement})$ is
    projected on this point.
  \end{enumerate}
\end{proposition}
\begin{sketchofproof}
  Consider a principal element~$\principalelement$
 of~$\InfGen{\delta}$ and its defining preprincipal
 element~$\preprincipalelement$ such that~${\InfGen{\delta}
 \preprincipalelement = \mu \preprincipalelement}$
 (resp.~${\InfGen{\delta} \preprincipalelement \not = 0}$
 and~${\InfGen{\delta}^{2} \preprincipalelement = 0}$). Then, the
 relation~${\projection{\InfGen{\delta}}{\principalelement}
 \preprincipalelement = \preprincipalelement
 \ExponentialMap{-\log(\preprincipalelement)} }$
 (resp.~${\projection{\InfGen{\delta}}{\principalelement}
 \preprincipalelement = 0}$) holds and
 thus~${\projection{\InfGen{\delta}}{\principalelement}
 \preprincipalelement}$ is equal to~$1$ (resp.~$0$) (to be more
 precise~${\partial \projection{\InfGen{\delta}}{\principalelement}
 \preprincipalelement /\partial \variable}$ is equal to~$0$ for
 all~$\variable$ in~$\variableset$ and
 thus~$\projection{\InfGen{\delta}}{\principalelement}
 \preprincipalelement$ is in~$\basefield$).  If we
 denotes~${\preprincipalelement -1}$ (resp.~$\preprincipalelement$)
 by~$\slice_{\principalelement}$, the
 ideal~$\slice_{\principalelement}\localization{\basefield[\variableset]}{\multiplicativeset}$
 is include in~$\ker\,
 \projection{\InfGen{\delta}}{\principalelement}$.  Furthermore, as we
 suppose that the flow~$\ExponentialMap{\flowtime\InfGen{\delta}}$
 acts regularly, its orbits have the same dimension~$1$ and thus, the
 associated invariant coordinates
 ring~$\basefield[\invariantset]$
 is of dimension~${\variablenb-1}$. Hence, the quotient
 algebra~$\localization{\basefield[\variableset]}{\multiplicativeset}/
 \slice_{\principalelement} \localization{\basefield[\variableset]}
 {\multiplicativeset}$ allows to describe
 algebraically~$\basefield[\invariantset]$. Let us now, see some
 properties of the associated algebraic
 variety~$V(\slice_{\principalelement})$.  The first above assertion
 is just the definition of a projection
 on~$V(\slice_{\principalelement})$.  \par\noindent 2) Remark that as
 the flow~$\ExponentialMap{\flowtime\InfGen{\delta}}$ is an
 homomorphism and relation~${\ExponentialMap{\flowtime\InfGen{\delta}}
 \InfGen{\delta} =
 \InfGen{\delta}\ExponentialMap{\flowtime\InfGen{\delta}} }$ holds, we
 have~${ \principalelement( \ExponentialMap{\flowtime\InfGen{\delta}}
 \variableset )^{\indicep} (\InfGen{\delta}^{\indicep}
 \ExponentialMap{\flowtime\InfGen{\delta}}Z) = \principalelement
 (\ExponentialMap{\flowtime\InfGen{\delta}}Z)^{\indicep}
 (\ExponentialMap{\flowtime\InfGen{\delta}} \InfGen{\delta}^{\indicep}
 Z) = \ExponentialMap{\flowtime\InfGen{\delta}} \left(
 \principalelement(\variableset)^{\indicep}
 \InfGen{\delta}^{\indicep}Z \right)}$, for all integer~$\indicep$. In
 order to show that the relation~${ \principalelement(
 \ExponentialMap{\flowtime\InfGen{\delta}} \variableset ) =
 \ExponentialMap{\flowtime\InfGen{\delta}} \principalelement(
 \variableset )}$ holds, remark that
 as~$\ExponentialMap{\flowtime\InfGen{\delta}}$ is a
 homomorphism,~$\preprincipalelement(
 \ExponentialMap{\flowtime\InfGen{\delta}} \variableset )$ is equal
 to~$\ExponentialMap{\flowtime\InfGen{\delta}}
 \preprincipalelement(\variableset)$.  If the principal
 element~$\principalelement$ is equal
 to~$\log(\preprincipalelement)/\eigenvalue$
 (resp.~${\preprincipalelement/\InfGen{\delta}\preprincipalelement}$)
 with~${\InfGen{\delta}\preprincipalelement = \eigenvalue
 \preprincipalelement}$
 (resp.~${\InfGen{\delta}\preprincipalelement\not = 0}$
 and~${\InfGen{\delta}^2\preprincipalelement=0}$) then the
 transcendental element~${\log(
 \ExponentialMap{\flowtime\InfGen{\delta}}\preprincipalelement
 )/\eigenvalue}$
 (resp.~${\ExponentialMap{\flowtime\InfGen{\delta}}(\preprincipalelement
 / \InfGen{\delta} \preprincipalelement)}$) is equal
 to~${\log(\preprincipalelement
 e^{\flowtime\eigenvalue})/\eigenvalue}$
 (resp.~${(\preprincipalelement +
 \flowtime\InfGen{\delta}\preprincipalelement) /
 \InfGen{\delta}(\preprincipalelement + \flowtime)}$) and
 thus,~$\principalelement( \ExponentialMap{\flowtime\InfGen{\delta}}
 \variableset)$ is equal to~${\principalelement+\flowtime}$ which is
 also equal
 to~$\ExponentialMap{\flowtime\InfGen{\delta}}\principalelement$.
 Above relations show that~$
 (\projection{\InfGen{\delta}}{\principalelement}
 \ExponentialMap{\flowtime\InfGen{\delta}}\variableset
 )(\originalsolution)$ is equal to~$(
 \ExponentialMap{\flowtime\InfGen{\delta}}
 \projection{\InfGen{\delta}}{\principalelement} \variableset
 )(\originalsolution)$.  As the flow leaves the image
 of~$\projection{\InfGen{\delta}}{\principalelement}$ invariant
 (i.e.~${\ExponentialMap{\flowtime\InfGen{\delta}}
 \projection{\InfGen{\delta}}{\principalelement} \variableset =
 \projection{\InfGen{\delta}}{\principalelement} \variableset}$), this
 quantity is equal to~$
 (\projection{\InfGen{\delta}}{\principalelement} \variableset
 )(\originalsolution)$. Hence, two points~$\originalsolution$
 and~$(\ExponentialMap{\flowtime\InfGen{\delta}} \variableset
 )(\originalsolution)$ in the same orbit
 of~$\ExponentialMap{\flowtime\InfGen{\delta}}$ are projected onto the
 same point of~$V(\slice_{\principalelement})$.  \par\noindent 3)
 If~$\inducedsolution$ is in~$V(\slice_{\principalelement})$
 then~$\slice_{\principalelement}(\inducedsolution)$ is equal to~$0$
 and thus,~$\principalelement(\inducedsolution)$ is also equal to~$0$.
 In that case, by
 construction~$(\projection{\InfGen{\delta}}{\principalelement}Z)(\inducedsolution)$
 is equal to~$\inducedsolution$.
\end{sketchofproof}
The orbits of~$\ExponentialMap{\flowtime\InfGen{\delta}}$ cross the
hyperplane~$V(\slice_{\principalelement})$ transversally. In the
sequel, we denote this hyperplane by~$\hyperplan$. Let us show now how
works our reduction process.
\subsection{Invariantization and Parameterization}
\label{sec:InvariantizationAndParameterization}
\subsubsection{Reduction of Algebraic Systems.}
Consider the variety~$V(F)$ in~$\basefield^{\variablenb}$ defined by
the ideal spanned by~$F$ in~$\localization{\basefield[\variableset]}
{\multiplicativeset}$ (for the sake of simplicity, we suppose that~$F$
is prime).
\paragraph{Parameterization of an algebraic variety invariant under
  the action of~$\ExponentialMap{\flowtime\InfGen{\delta}}$.}
If~$\InfGen{\delta}$ is an affine derivation such that the
relation~${\InfGen{\delta}F=\eigenvalue F}$ holds, it is an
infinitesimal generator of a Lie
symmetry~$\ExponentialMap{\flowtime\InfGen{\delta}}$ that leaves the
variety~$V(F)$ invariant as shown by the following relations holding
for all~$\inducedsolution$ in~$\basefield^{\variablenb}$,
\begin{equation}
  \label{eq:1}
  \textstyle
  \originalsolution := 
  \bigl(\ExponentialMap{\flowtime\InfGen{\delta}}\variableset\bigr)
  (\inducedsolution),\
  F(\originalsolution) 
  := \bigl(F\bigl(\ExponentialMap{\flowtime\InfGen{\delta}}\variableset\bigr)\!\bigr)
  (\inducedsolution)
  = \bigl(\ExponentialMap{\flowtime\InfGen{\delta}} F(\variableset)\bigr)
  (\inducedsolution)
  = e^{\flowtime\eigenvalue}F(\inducedsolution).
\end{equation}
As shown by Proposition~\ref{prop:CoordinatesChange}, the
hyperplane~$\hyperplan$ is a linear cross-section of the orbit
of~$\ExponentialMap{\flowtime\InfGen{\delta}}$ i.e.\ a variety that
intersects these orbits in a single point.  Furthermore, there is a
variety~$\inducedvariety$ induced by~$V(F)$ in~$\hyperplan$ such that
the variety~$V(F)$ is the image of~${\basefield \times
\inducedvariety}$ by the action of the
flow~$\ExponentialMap{\flowtime\InfGen{\delta}}$.  Let us described
now~$\inducedvariety$.
\paragraph{Invariantization of purely algebraic systems.}
The variety~$\inducedvariety$ is defined by the
intersection~${V(F\,\textup{mod}\, \slice_{\principalelement}) \cap
\hyperplan}$ (if~$\slice_{\principalelement}(\inducedsolution)$
and~$(F\,\textup{mod}\,\slice_{\principalelement})(\inducedsolution)$
are equal to zero, then the relations~$F(\inducedsolution)=0$ hold).
As~$\slice_{\principalelement}$ is linear, a description
of~$\inducedvariety$ is obtained by a simple substitution in the
equations describing~$V(F)$ (compare with the replacement invariant
studied in~\cite{HubertKogan2006}) as shown by the following example:
\begin{example}
  \label{ex:LadderAndBox}
  Let us consider the following purely algebraic system:
  \begin{equation}
    \label{eq:LadderAndBox}
    \OrdinaryDifferentialSystem:\qquad
    (y-b)^2+a^2 = l^2/4,\quad
    (x-a)^2+b^2 = l^2/4,\quad
    x^2+y^2 = l^2.
  \end{equation}
  Using results of Section~\ref{sec:PrincipalElementComputation}, we
  determine its expanded affine Lie symmetries:
  \begin{equation}
    \label{eq:LadderAndBoxInfGen}
    \textstyle
    \InfGen{\delta}_{1} :=
    x\frac{\partial\hfill}{\partial x}
    + y\frac{\partial\hfill}{\partial y}
    + a \frac{\partial\hfill}{\partial a}
    + b \frac{\partial\hfill}{\partial b}
    + l \frac{\partial\hfill}{\partial l},
    \ 
    \InfGen{\delta}_{2} :=
    -y\frac{\partial\hfill}{\partial x}
    + x\frac{\partial\hfill}{\partial y}
    + (b - y) \frac{\partial\hfill}{\partial a}
    + (x - a) \frac{\partial\hfill}{\partial b}.
  \end{equation}
  As~$l$ is a preprincipal element of~$\InfGen{\delta}_{1}$, the
  solutions~$\variableset$ of system~(\ref{eq:LadderAndBox}) are
  represented by the parameterization~${\variableset =
    \ExponentialMap{\flowtime \InfGen{\delta}_{1}}
    \invariant{\variableset}_{1}}$
  where~$\invariant{\variableset}_{1}$ are the solutions of an
  invariant system obtained by the intersection
  of~(\ref{eq:LadderAndBox}) with the hyperplane~${l=1}$:
  \begin{equation}
    \label{eq:LadderAndBox1}
    \forall \variable \in \{x,y,a,b,l\},\
    \begin{array}[t]{ccc}
      \variable &=& e^{\flowtime} \invariant{\variable}_{1}, \\
      \invariant{l}_{1} &=& 1,
    \end{array}
    \
    \left\lbrace
      \begin{array}{ccc}
        (\invariant{y}_{1}-\invariant{b}_{1})^2+{\invariant{a}_{1}}^{\! 2} &=& 1/4,\\
        (\invariant{x}_{1}-\invariant{a}_{1})^2+{\invariant{b}_{1}}^{\! 2} &=& 1/4,\\
        {\invariant{x}_{1}}^{\! 2} + {\invariant{y}_{1}}^{\! 2} &=& 1.
      \end{array}
    \right.
  \end{equation}
  As~$\InfGen{\delta}_{1}$ and~$\InfGen{\delta}_{2}$ form an abelian
  Lie algebra, this last derivation is an infinitesimal generators of
  a Lie symmetry of~$\InfGen{\delta}_{1}$; it could be used to reduce
  further the system~(\ref{eq:LadderAndBox1}) (the Lie algebra spanned
  by~$\InfGen{\delta}_{1}$ and~$\InfGen{\delta}_{2}$ is abelian and
  thus solvable).  In fact, the linear form~${a-I b-x}$ is a
  preprincipal element of~$\InfGen{\delta}_{2}$ associated to the
  principal element~$\log({a-I b-x})/I$.  Using
  symmetry~$\InfGen{\delta}_{2}$, solutions of
  system~(\ref{eq:LadderAndBox1}) could be represented as follow:
  \begin{equation}
    \label{eq:LadderAndBox2}
    \forall\variable_{1}  \in 
    \{\invariant{x}_{1},\invariant{y}_{1},\invariant{a}_{1},\invariant{b}_{1}\},\
    \begin{array}[t]{ccc}
      \invariant{\variable}_{1} &=&
      \ExponentialMap{\flowtime\InfGen{\delta}_{2}}
      \invariant{\variable}_{2},  \\
      \invariant{x}_{2} &=& \invariant{a}_{2}-I \invariant{b}_{2} - 1,
    \end{array}
    \
    \left\lbrace
      \begin{array}{ccc}
        (\invariant{y}_{2}-\invariant{b}_{2})^2+{\invariant{a}_{2}}^{\! 2}&=&1/4,
        \\
        (1 + I \invariant{b}_{2})^2+{\invariant{b}_{2}}^{\! 2} &=& 1/4, \\
        {(\invariant{a}_{2} - I \invariant{b}_{2} - 1)}^{\! 2} 
        + {\invariant{y}_{2}}^{\! 2} &=& 1.
      \end{array}
    \right.
  \end{equation}
  In this particular example, the positive dimensional
  system~(\ref{eq:LadderAndBox}) is represented
  \begin{itemize}
  \item by a zero-dimensional algebraic
    system~(\ref{eq:LadderAndBox2}) that furnishes initial
    values~$\invariant{Z}_{2}$
  \item to an explicit linear differential system whose solutions
    are~${ \invariant{\variableset}_{1} =
      \ExponentialMap{\flowtime\InfGen{\delta}_{2}}
      \invariant{\variableset}_{2}}$ (this system associated to the
    derivation~$\InfGen{\delta}_{2}$ is simple enough to be explicitly
    solved in closed form but in more complicated cases it could also
    be considered as a black box representation solved by purely
    numerical methods);
  \item these values~$\invariant{\variableset}_{1}$ constitute an
    initial condition set of the linear differential system---induced
    by the derivation~$\InfGen{\delta}_{1}$---such that resulting
    solutions~$\variableset$ parameterize the variety defined by
    system~(\ref{eq:LadderAndBox2}).
  \end{itemize}
\end{example}
We show now that the same type of results exists for differential
systems.
\subsubsection{Reduction of Differential systems.}
\begin{hypotheses}
  \emph{Restriction on Symmetries Specific to Differential Case } ---
  Lie symmetries of a given vector field~$\InfGen{D}$ acting only on
  its state variables~(${\dot{z}\not =0}$) could be used for Lie based
  integration but not for the previous reduction process.  Thus, we
  suppose that~$\InfGen{D}$ have an expanded Lie symmetry that acts at
  least on one of its parameter~(${\InfGen{D}z=0}$) and that there is
  an associated principal elements such that the associated linear
  form~$\slice_{\principalelement}$ satisfies the
  relation~${\InfGen{D}\slice_{\principalelement}=0}$.
\end{hypotheses}
\paragraph{Invariantization of an infinitesimal generator.}
Given any derivation~$\InfGen{D}$ acting
on~$\basefield[\variableset]$, the
sequence~(\ref{eq:AffineExactSequence}) induces a
derivation~$\overline{\InfGen{D}}$ acting
on~$\basefield[\invariantset]$ such that the
relation~${\projection{\InfGen{\delta}}{\principalelement} \circ
  \InfGen{D} = \overline{\InfGen{D}}\circ
  \projection{\InfGen{\delta}}{\principalelement}}$ holds.  The
exponentiation of~$\overline{\InfGen{D}}$ (see
Definitions~\ref{def:FormalFlowInfinitesimalGenerator}) induces a
flow~$\ExponentialMap{\flowtime\overline{\InfGen{D}}}$ on the
hyperplane~$\hyperplan$ that is the invariantization of the
flow~$\ExponentialMap{\flowtime\InfGen{D}}$ acting
on~$\basefield^{\variablenb}$.  Under above hypotheses, the
flow~$\ExponentialMap{\flowtime\overline{\InfGen{D}}}$ is just the
restriction of~$\ExponentialMap{\flowtime\InfGen{D}}$ on~$\hyperplan$;
in fact, as~${\InfGen{D}\slice_{\principalelement}=0}$ the
relation~${\ExponentialMap{\flowtime\InfGen{D}}\slice_{\principalelement}
  = \slice_{\principalelement}}$ holds and thus, the
flow~$\ExponentialMap{\flowtime\InfGen{D}}$ maps any point
of~$\hyperplan$ to another point of this hyperplane. The set of orbits
of~$\ExponentialMap{\flowtime\InfGen{D}}$
in~$\basefield^{\variablenb}$ is projected onto the set of orbits
of~$\ExponentialMap{\flowtime\InfGen{D}}$ in~$\hyperplan$.  Let us see
now the condition on~$\InfGen{\delta}$ and~$\InfGen{D}$ that allows to
parameterize the set of orbits
of~$\ExponentialMap{\flowtime\InfGen{D}}$
in~$\basefield^{\variablenb}$ by the set of orbits
of~$\ExponentialMap{\flowtime\InfGen{D}}$ in~$\hyperplan$ and the
map~$\ExponentialMap{\flowtime\InfGen{\delta}}$.
\paragraph{Parameterization of vector field~$\InfGen{D}$ invariant
  under the action of the
  flow~$\ExponentialMap{\flowtime\InfGen{\delta}}$.}
If~$\InfGen{\delta}$ is the infinitesimal generator of a symmetry of
derivation~$\InfGen{D}$, according to
Definition~\ref{def:LiePointSymmetry}, the relation~${[\InfGen{D},
  \InfGen{\delta}]= \lambda \InfGen{D}}$ holds. The Baker Campbell
Hausdorff formula (Lemma~\ref{lem:BCH}) shows that the
relation~${\ExponentialMap{\flowtime_{1}\InfGen{\delta}}
  \ExponentialMap{(1+\flowtime_{1}\lambda)\flowtime_{2}\InfGen{D}}
  \ExponentialMap{-\flowtime_{1}\InfGen{\delta}} =
  \ExponentialMap{\flowtime_{2}\InfGen{D}}}$ holds.  This implies that
any orbit of~$\ExponentialMap{\flowtime\InfGen{D}}$
in~$\basefield^{\variablenb}$ is the image of an orbit
of~$\ExponentialMap{\flowtime\InfGen{D}}$ in~$\hyperplan$ by the
flow~$\ExponentialMap{\flowtime\InfGen{\delta}}$. Let us explicit all the process described above through an example.
\begin{example} \label{ex:FitzHughNagumo} Consider a FitzHugh Nagumo
  model (see~\S~7 in~\cite{Murray2002}):
  \begin{equation}
    \label{eq:FitzHughNagumoModel}
    \dot{a}=\dot{b}=\dot{c}=\dot{d}=0,\quad
    \dot{x}= (x-x^3/3 - y + d)c,\quad
    \dot{y}/\textup{d}t = (x+ a - b y)/c.
  \end{equation}
  The derivation~${ \InfGen{\delta} := \lfrac{\partial }{\partial y} +
    b \lfrac{\partial }{\partial a} + \lfrac{\partial }{\partial d} }$
  is an infinitesimal generator of the following one-parameter
  group~${{y \rightarrow y + \lambda},\ {a \rightarrow a + b\lambda},\
    {d \rightarrow d + \lambda}}$ that is composed of symmetries of
  the system~(\ref{eq:FitzHughNagumoModel}). As the
  relation~${\InfGen{\delta} d =1}$ holds,~$d$ is a (pre)principal
  element of~$\InfGen{\delta}$ and the
  solutions~${\variableset:=\{x,y,a,b,c,d\}}$ of
  system~(\ref{eq:FitzHughNagumoModel}) are described by the
  parameterizations~${\variableset = \ExponentialMap{d\InfGen{\delta}}
    \invariant{\variableset}}$ where~$\invariant{\variableset}$ are
  solutions of a differential system on the hyperplane~$V(d)$;
  hence,~$\variableset$ are given by the equations:
  \begin{equation}
    \label{eq:FitzHughNagumoModel2}
    y = \invariant{y} + d,\ a = \invariant{a} -b d,\quad
    \dot{x} =\bigl( x-x^3/3 - \invariant{y} \bigr) c,\
    \dot{\invariant{y}} = (x+ \invariant{a} - b \invariant{y})/c.
  \end{equation}
\end{example}
\section{Conclusion}
\label{sec:Conclusion}
In this note, we consider the computation of affine expanded Lie
symmetries of a given algebraic system and show how this system could
be rewrite in an invariant coordinates set for these symmetries in
order to reduce the number of involved parameters. As this process is
based on the computation of Jordan normal form and numerical linear
algebra, its complexity is quasi-polynomial in input's size and likely
polynomial for the great majority of practical cases.
\paragraph{Extension of the reduction process to more general types of
  derivations.}  The manipulation presented in previous sections for
affine derivations could be used for non-affine symmetries that occurs
in practice as shown below.
\begin{example}
  \label{ex:NonLinearInfGen}
  Let us consider the following algebraic system:
  \begin{equation}
    \label{eq:NonLinearInfGen_Sys}
    {1} - {x} +{x}^2{y} = 0,\qquad
    {b} -    {x}^2{y} = 0.
  \end{equation}
  One can check that the following infinitesimal generator
  \begin{equation}
    \label{eq:NonLinearInfGen_InfGen}
    \textstyle
    \InfGen{\delta} := 
    x^{2}\lfrac{\partial\hfill}{\partial x}
    + (1 - 2 x y)\lfrac{\partial\hfill}{\partial y}
    + x^{2}\lfrac{\partial\hfill}{\partial b} 
  \end{equation}
  is associated to the following one-parameter group of automorphisms:
  \begin{equation}
    \label{eq:NonLinearInfGen_Sym}
    \textstyle
    \ExponentialMap{\lambda\InfGen{\delta}}:
    x \to \frac{x}{1-x\lambda},\quad
    y \to 
    \left(y+  \frac{\lambda}{1-x\lambda}\right)(x\lambda-1)^2,\quad
    b  \to  b + x \frac{x\lambda}{1-x\lambda},
  \end{equation}
  that is a one-parameter group of Lie point symmetries of the
  system~(\ref{eq:NonLinearInfGen_Sys}).  Remark
  that~${\principalelement:=b/(x(x-b))}$ is a principal element
  of~(\ref{eq:NonLinearInfGen_InfGen})
  (${\InfGen{\delta}\principalelement=1}$) and thus, all that we have
  done previously could be repeated i.e.\ the invariant coordinate
  set:
  \begin{equation}
    \label{eq:NonLinearInfGen_Invariant}
    \projection{\InfGen{\delta}}{\principalelement} x = x-b,
    \quad
    \projection{\InfGen{\delta}}{\principalelement} y = 
    \lfrac{(y x^2-b)}{(b-x)^2},
    \quad
    \projection{\InfGen{\delta}}{\principalelement} b = 0,
  \end{equation}
  allows to represent the solutions set~$(x,y)$
  of~(\ref{eq:NonLinearInfGen_Sys}) as follow
  \begin{equation}
    \label{eq:NonLinearInfGen_InvariantSys}
    x = \invariant{x}+b,\ y =(\invariant{y} \invariant{x}^{2}+b) /(\invariant{x} + b)^2
    \quad
    \forall (\invariant{x},\invariant{y})\ \textrm{s.t.}\
    {1} - {\invariant{x}} +{\invariant{x}}^{2}{\invariant{y}} = 0,\
    {\invariant{x}}^{2}{\invariant{y}} = 0.
  \end{equation}
\end{example}
The results presented here could likely be extended for more general
types of derivations but we do not know if the associated computations
are feasible.
\subsubsection*{Acknowledgments.}
The author is grateful to G.\ Renault, \'E.\ Schost, and M.\
Safey\,El\,Din for many pleasant and useful discussions related to
this note.
 \bibliographystyle{splncs} \bibliography{Sedoglavic2006}
\end{document}